\documentclass[
    ,final            
  ]
  {aipproc}
\layoutstyle{6x9}
\usepackage{subfigure}
\usepackage{mathrsfs}
\usepackage{graphicx}
\usepackage{times}
\usepackage{pgf,pgfarrows,pgfnodes,pgfautomata,pgfheaps,pgfshade}
\usepackage{amsmath,amssymb}
\usepackage{fix-cm}
\usepackage{geometry}
\usepackage[cmtip,arrow]{xy}
\usepackage{pb-diagram,pb-xy}
\usepackage{rotating}

\newcommand{\beq}{\begin{equation}}
\newcommand{\eeq}{\end{equation}}

\input{aipcheck}

\begin{document}

\title{MOND as the weak-field limit of an extended metric theory of gravity
}

\author{S.~Mendoza}{address = {Instituto de Astronom\'{\i}a,
Universidad Nacional Aut\'onoma de M\'exico, AP 70-264, Distrito
Federal 04510, M\'exico.} } 
\author{T. Bernal}{address = {Instituto de Astronom\'{\i}a,
Universidad Nacional Aut\'onoma de M\'exico, AP 70-264, Distrito
Federal 04510, M\'exico.} } 
\author{J.~C.~Hidalgo}{address = {Instituto de Astronom\'{\i}a,
Universidad Nacional Aut\'onoma de M\'exico, AP 70-264, Distrito
Federal 04510, M\'exico.} } 
\author{S. Capozziello}{address = { Dipartimento di Scienze Fisiche,
Universit\`a degli Studi di Napoli ``Federico II'', 
Complesso Universitario di Monte Sant'Angelo, Edificio N, via Cinthia,
80126, Napoli, Italy \\
I.N.F.N. - Sezione di Napoli, Complesso Universitario di Monte Sant'Angelo, 
Edificio G, via Cinthia, 80126, Napoli, Italy} } 
\date{\today}

\begin{abstract}
  We show that the Modified Newtonian Dynamics (MOND) regime can be
fully recovered as the weak-field limit of a particular theory
of gravity formulated in the metric approach.  This is possible when
Milgrom's acceleration constant is taken as a fundamental
quantity which couples to the theory in a very consistent manner.  
As  a consequence, the scale invariance of the gravitational
interaction is naturally broken.  In this sense,  Newtonian gravity is
the weak-field limit of general relativity and MOND is
the weak-field limit of that particular extended theory of 
gravity. 
\end{abstract}
\maketitle


\section{Extended relativistic gravity}

  As explained by the presentations of Hidalgo and Bernal in this
conference series, the inclusion of Milgrom's acceleration constant \( a_0
\) as a fundamental quantity in a description of gravity leads to a great
deal of understanding in a wide variety of non-relativistic 
astrophysical systems.   As such, let us assume that a point mass \( M \) 
located at the origin
of coordinates generates a relativistic gravitational field in the MONDian
regime and that a metric formalism describes the field equations.
This problem is characterised by the
speed of light in vacuum \( c \), the mass \( M \) of the central object
generating the gravitational field, Newton's constant of gravity \( G \)
and Milgrom's acceleration constant \( a_0 \).  With these parameters,
two ``fundamental lengths'' can be built:

\begin{equation}
  r_\text{g} :=  G M / c^2 , \qquad l_M := \left(  G M / a_0
    \right)^{1/2},
\label{eq01}
\end{equation}

\noindent where \( r_\text{g} \) is the gravitational radius and \(
l_M \) is the mass-length scale as described in \cite{mendoza11} and
by Hidalgo's presentation in this conference series.  When relativistic
effects are taken into account for the gravitational field, then standard
general relativity should be recovered in the limit  \( l_M / r \gg 1
\), and a relativistic version of MOND should be obtained when \( l_M /
r \ll 1 \).  This shows that the pursue of a complete metric description
leads one to consider the scale-dependence of gravity.

  The length scales presented in equation~\eqref{eq01} must somehow
appear in a relativistic theory of gravity which includes the fundamental
nature of the constant \( a_0 \).  For example, in the metric formalism,
a generalised Hilbert action \( S_\text{H} \) can be written in the
following way:

\begin{equation}
   S_\text{H}  = - \frac{ c^3 }{ 16 \pi G L_M^2 } \int{ f(\chi) \sqrt{-g}
     \, \mathrm{d}^4x},
\label{eq03}
\end{equation}

\noindent which slightly differs from its traditional form
(see e.g. \cite{capozziellobook}) since we have introduced  the
dimensionless Ricci's scalar \( \chi := L_M^2 R \), where \( R \) is
the standard Ricci's scalar and \( L_M \) defines a length fixed by the
parameters of the theory.  The explicit form of the length \( L_M \)
has to be obtained once a certain known limit of the theory is taken,
usually a non-relativistic limit.  In fact, since equation~\eqref{eq01}
defines two natural lengths of the theory it is coherent to postulate  the
following relation between all these lengths:

\begin{equation}
  L_M \propto  r_\text{g}^\alpha l_M^\beta, \qquad \text{with} \qquad
    \alpha + \beta = 1.
\label{eqlm}
\end{equation}

\noindent Note that the definition of \( \chi
\) gives a correct dimensional character to the action~\eqref{eq03},
something that is not completely clear in all previous works dealing
with a metric description of the gravitational field.  For \( f(\chi)
= \chi \), the standard Einstein-Hilbert action is obtained. Using the
usual form for the matter action \( S_\text{m} = - \left( 1 / 2 c \right) 
\int{ {\cal L}_\text{m} \, \sqrt{-g} \, \mathrm{d}^4x } \), with a matter
Lagrangian \( {\cal L}_\text{m} \),  then the
null variations of the complete action \( \delta \left( S_\text{H} +
S_\text{m} \right) = 0 \) yield the following field equations:

\begin{equation}
    f'(\chi) \, \chi_{\mu\nu} - \frac{ 1 }{ 2 } f(\chi) g_{\mu\nu} - L_M^2 
      \left( \nabla_\mu \nabla_\nu -g_{\mu\nu} \Delta \right) f'(\chi)
    = \frac{ 8 \pi G L_M^2 }{ c^4} T_{\mu\nu},
\label{eq06}
\end{equation}

\noindent where the dimensionless Ricci tensor \( \chi_{\mu\nu}:= 
L_M^2 R_{\mu\nu} \) and \( R _{\mu\nu} \) is the standard Ricci tensor.
The Laplace-Beltrami operator has been written as \( \Delta :=
\nabla^\alpha \nabla_\alpha \) and the prime denotes derivative with
respect to its argument.  The energy-momentum tensor \( T_{\mu\nu} \)
is defined through the following standard relation: \( \delta S_\text{m}
= - \left( 1 / 2 c \right) T_{\alpha\beta} \, \delta g^{\alpha\beta} \).
In here and in what follows, we choose a (\(+,-,-,-\)) signature for
the metric \( g_{\mu\nu} \) and use Einstein's summation convention over
repeated indices.

  The trace of equation~\eqref{eq06} is:
\begin{equation}
  f'(\chi) \, \chi  - 2 f(\chi) + 3 L_M^2  \, \Delta  f'(\chi) = 
    \frac{ 8 \pi G L_M^2 }{ c^4} T,
\label{eq08}
\end{equation}

\noindent where \( T := T^\alpha_\alpha \).

\section{MONDian weak-field limit}

  Under the assumption of a power-law relation for the function \( f(\chi)
\), i.e.

\begin{equation}
  f(\chi) = \chi^b
\label{eq11}
\end{equation}

\noindent it follows that the trace~\eqref{eq08} can be approximated to
order of magnitude since \( \mathrm{d} / \mathrm{d} \chi \approx 1 /
\chi \), \( \Delta \approx - 1 / r^2  \) and the mass density \( \rho
\approx  M / r^3 \) yielding:

\begin{equation}
   \chi^b  \left( b - 2 \right) - 3 b L_M^2  \frac{ \chi^{(b-1)} }{ r^2 }
     \approx \frac{ 8 \pi G M L_M^2 }{ c^2 r^3}.
\label{eq12}
\end{equation}

Note that the second term on the left-hand side of equation~\eqref{eq12}
is much greater than the first term when 
\( R  r^2  \lesssim 3 b / \left(2 - b \right) \) and so, since 
\( R \approx \kappa = R_\text{c}^{-2} \), where \( \kappa \) is the Gaussian
curvature of space and \( R_\text{c} \) its radius of curvature, then
\(  R_c \gg r \).  This should occur in the weak-field regime, 
where MONDian effects are expected.  For a metric description of gravity, 
this limit must correspond to the relativistic regime of MOND. 
In what follows  we will only deal with this approximation and so, 
equation~\eqref{eq12} takes the following form:

\begin{equation}
    R^{ (b-1)} \approx - \frac{ 8 \pi G M  }{ 3 b c^2 r
      L_M^{2 \left( b - 1 \right) } }.
\label{eq15}
\end{equation}

  Using the fact that at second order of approximation \(  R =  - \left(
2 / c^2 \right) \nabla^2 \phi = +\left( 2 / c^2 \right) \nabla \cdot
\bold{a}, \) for a non-relativistic potential \( \phi \) and an
acceleration \( \bold{a} \), it follows that

\begin{equation}
  a \approx - c^{\left( 2 b- 4  \right)/\left( b - 1 \right) } r^{ \left( b -
    2 \right) / \left( b - 1 \right) } L_M^{-2} \left( G M  \right)^{
    1 / \left( b - 1 \right) }, 
\label{eq19}
\end{equation}

\noindent which converges to a MOND-like acceleration regime (i.e. \( a
\propto 1 / r \)) if \( b = 3/2 \).  At the lowest order of approximation a
description of the gravitational force should not include the velocity of
light and so \( L_M \propto 1/c \).  Combining this result with
equation~\eqref{eqlm} then relation~\eqref{eq19}  means that

\begin{equation}
  a \approx - \frac{ \left( a_0 G M \right)^{1/2} }{ r },
\label{eq23}
\end{equation}

\noindent which is the traditional form of MOND in spherical symmetry.

  As explained by \cite{bernal11} in a more rigorous form, the
result obtained in equation~\eqref{eq23} is of general character, since it
can be formally proved at the lowest order of perturbation for a \( f(\chi) =
\chi^{3/2} \) metric theory of gravity far away from the mass sources, i.e.
when \( R_\text{c} \gg r \), or equivalently when \( l_M \gg r_\text{g} \).
In \cite{bernal11} it is also discussed the Noether symmetries of the
problem and for this particular case the conserved charge of the problem is
proportional to \( r_\text{g}^2 l_M \).

  The metric theory of gravity presented here is by no means a complete
description at all scales of gravitation.  It only deals with the MONDian
regime of gravity, i.e.  when \( l_M \gg r_\text{g} \) is valid. In other
words, our description breaks the scale invariance of gravity in a more
general way than the one described in \cite{mendoza11}.

The mass dependence of \( \chi \) means that the mass needs to appear
on Hilbert's action~\eqref{eq03}.  This is traditionally not the case,
since that action is thought to be purely a function of the geometry
of space-time due to the presence of mass and energy.  However, it was
\cite{sobouti06} who first encountered this peculiarity in the  Hilbert
action when dealing with a metric generalisation of MOND and later on
also discussed by~\cite{mendoza07}.  Following the remarks of \cite{sobouti06},
one should not be surprised if some of the commonly accepted notions,
even at the fundamental level of the action, require generalisations
and re-thinking.  An extended metric theory of gravity goes beyond the
traditional general relativity ideas and in this way, we probably need
to change our standard view of its fundamental principles.  

  With these ideas in mind it is interesting to note that recently 
\cite{harko11} have proposed a modified \( F(R,T) \) 
theory of gravity, where \( T \) represents the trace of the
energy-momentum tensor.  The theory developed in this article is a particular
example of their proposal with the identification \( F(R,T) = f(\chi) /
L_M^2 \).  We are developing a cosmological theory with these ideas in mind
and we are also applying this full theory to gravitational lenses and to the
dynamics of clusters of galaxies.  The obtained results will be reported
elsewhere.

\section{Acknowledgements}

This work was supported by a DGAPA-UNAM grant (PAPIIT IN116210-3).
The authors TB, JCH \& SM  acknowledge economic support from CONACyT:
207529, 51009, 26344.

\bibliographystyle{plain}

\end{document}